\documentclass[aps,nofootinbib,notitlepage,superscriptaddress]{revtex4}
\usepackage{graphicx}
\usepackage{dcolumn}
\usepackage{bm}
\usepackage{amsmath,amssymb}
\usepackage{float}
\usepackage{multirow}
\usepackage{slashed}
\usepackage{xcolor}
\usepackage{physics}
\usepackage{multirow}
\usepackage[colorlinks=true, pdfstartview=FitV, bookmarks=true, bookmarksnumbered=true, breaklinks]{hyperref}
\usepackage{mathtools,braket}
\usepackage{soul}
\usepackage{lipsum}  
\usepackage{color}
\definecolor{blue}{rgb}{0.0, 0.0, 1.0}
\definecolor{red}{rgb}{1.0, 0.0, 0.0}
\definecolor{royalblue}{rgb}{0.0, 0.14, 0.4}
\hypersetup{linkcolor=royalblue, citecolor=blue, urlcolor=royalblue}

\usepackage{hyperref}
\hypersetup{colorlinks=true,citecolor=blue,linkcolor=blue,urlcolor=blue}

\usepackage[mathlines]{lineno}
\usepackage{stackengine}

\usepackage{tikz,xcolor,hyperref}
\definecolor{lime}{HTML}{A6CE39}
\DeclareRobustCommand{\orcidicon}{%
	\begin{tikzpicture}
	\draw[lime, fill=lime] (0,0) 
	circle [radius=0.16] 
	node[white] {{\fontfamily{qag}\selectfont \tiny ID}};
	\draw[white, fill=white] (-0.0625,0.095) 
	circle [radius=0.007];
	\end{tikzpicture}
	\hspace{-2mm}
}

\foreach \x in {A, ..., Z}{%
	\expandafter\xdef\csname orcid\x\endcsname{\noexpand\href{https://orcid.org/\csname orcidauthor\x\endcsname}{\noexpand\orcidicon}}
}

\begin{document}
\title{Updated analyses of gluon distribution functions for the pion and kaon\\ from the gauge-invariant nonlocal chiral quark model }
\author{Parada.~T.~P.~Hutauruk\orcidA{}}
\email{phutauruk@pknu.ac.kr}
\affiliation{Department of Physics, Pukyong National University (PKNU), Busan 48513, Korea}
\affiliation{Department of Physics Education, Daegu University, Gyeongsan 39453, Korea}
\author{Seung-il Nam\orcidB{}}
\email{sinam@pknu.ac.kr}
\affiliation{Department of Physics, Pukyong National University (PKNU), Busan 48513, Korea}
\affiliation{Center for Extreme Nuclear Matters (CENuM), Korea University, Seoul 02841, Korea}
\affiliation{Asia Pacific Center for Theoretical Physics (APCTP), Pohang 37673, Korea}
\date{\today}
\begin{abstract}
In this work, we investigate the gluon distribution functions for the pion and kaon, in addition to the improved result of the valence-quark ones, in the gauge-invariant nonlocal chiral-quark model (NL$\chi$QM), in which the momentum dependence of the quark interactions is properly taken into account. We then analyze the gluon distribution functions, generated dynamically through the splitting functions in the DGLAP QCD evolution. By comparing with the recent lattice QCD results and JAM global analyses, it is found that the present numerical results for the gluon parton distribution functions for the pion exhibit a remarkable agreement, followed by the valence up-quark distribution results for the pion by reproducing the reanalyzed experimental data. Our prediction on the gluon distribution functions for the kaon is also consistent with the recent lattice data for the kaon within the errors.
\end{abstract}

\maketitle

\section{Introduction} \label{sec:intro}
The parton distribution function (PDF) is one of the excellent instruments to access the nonperturbative aspects of the hadron structure, besides the elastic form factor (EFF), transverse momentum dependent (TMD), parton distribution amplitude (PDA), fragmentation function (FF), and generalized parton distribution (GPD), and so forth~\cite{Berger:1979du}. Studying the PDF is very crucial to characterize the structure of the pion and kaon, and to further understand dynamical chiral symmetry breaking (DCSB), which is one of the features of nonperturbative QCD. In fact, our knowledge and understanding of the quark and gluon distribution functions (QDFs and GDFs) are insufficient compared to the nucleon ones, in particular for the GDFs, because of the lack of the meson targets in the experiment. Nowadays, the situation becomes worse because of the current controversy on the pion PDF's power law behavior at large-$x$, which shows different predictions and/or interpretations obtained among the theoretical models and analyses, once they are compared with the existing experimental data through the Drell-Yan process~\cite{Conway:1989fs}. Therefore, more collections of data and theoretical studies are really necessary to resolve the current controversy and to understand the quarks and gluons dynamics inside the light mesons.

Recently, a few suggestions to access the pion and kaon PDFs, as well as the corresponding form factors, through the Sullivan process~\cite{Sullivan:1971kd} in electron-ion collider (EIC)~\cite{Arrington:2021biu}, electron-ion collider in China (EicC)~\cite{Anderle:2021wcy} have been intensively discussed in the literatures~\cite{Chavez:2021koz,Arrington:2021biu,Anderle:2021wcy}. For example, in accessing the pion-EFF data, the Sullivan process has provided a significantly larger value of momentum transfer $Q^2$ coverage~\cite{JeffersonLab:2008jve}. Analogous to the pion EFF, they argued that accessing the PDF also becomes feasible in the Sullivan process~\cite{AbdulKhalek:2021gbh}. This process is somewhat different from the previous reaction process used to extract the pion PDF, which was mostly taken from the pion-induced Drell-Yan and $J/\psi$ production processes to access the pion GDF data. Also, the COMPASS++/AMBER experiment at CERN~\cite{Adams:2018pwt} has been proposed to measure the pion-nucleus Drell-Yan process cross-section. This will allow us to access more data in the large-$x$ region and the pion GDF. An accessible source of the kaon beam would allow us to collect data for the kaon PDFs. Note that the pion and kaon GDFs are one of the focus programs of future experiments of EIC~\cite{Arrington:2021biu}, EicC~\cite{Anderle:2021wcy}, and COMPASS++/AMBER~\cite{Adams:2018pwt}, therefore the study of the present work will be relevant.

Besides those future experiments, several theoretical studies and efforts have also been made to analyze the pion and kaon PDFs~\cite{Hutauruk:2016sug,Hutauruk:2018zfk,Jia:2018ary,Kock:2020frx,Nam:2012vm,Hutauruk:2018qku,Hutauruk:2021kej,Hutauruk:2019ipp,Cui:2021mom,Albino:2022gzs,dePaula:2022pcb,Bourrely:2022mjf}, as well as lattice QCD~\cite{Fan:2021bcr,Salas-Chavira:2021wui} and global analyses~\cite{Novikov:2020snp,JeffersonLabAngularMomentumJAM:2022aix,Barry:2021osv}, to understand the relevant dynamics of the quarks and gluons~\cite{Aguilar:2020uqw} inside the kaon and pion. Gluon dynamics are expected to be closely related to the confinement and gluon saturation at small-x~\cite{Pennington:2011xs,Armesto:2022mxy}, and are very challenging and exciting topics in nonperturbative QCD. However, these topics are out of the scope of the present work, but they absolutely deserve further investigation and study for future work. Many impressive signs of progress have been made so far in understanding the gluon momentum distributions in the kaon and pion. Here, we emphasize again that more theoretical studies with various approaches to studying the gluon distribution are certainly required to support the experimental physics programs since the gluon contribution to the pion and kaon masses are significant, and produces about $(30-40)$\% of their masses~\cite{Barry:2021osv}.

In this work, we first time investigate the pion and kaon PDFs in the framework of the gauge-invariant nonlocal chiral quark model (NL$\chi$QM), taking the momentum dependence properly into account. However, in the present work, we will concentrate on the pion and kaon GDFs. The NL$\chi$QM has been widely applied to compute the QDFs~\cite{Nam:2012vm}, quasi-parton distribution amplitudes (QPDA)~\cite{Nam:2017gzm}, and EFFs for the pion and kaon~\cite{Yang:2015avi}. In computing the pion and kaon GDFs, we employ the next-to-leading order Dokshitzer–Gribov–Lipatov–Altarelli–Parisi (NLO DGLAP) QCD evolution~\cite{Miyama:1995bd} to dynamically generate the GDFns at a specific factorization scale value of $Q^2$, which is chosen based on experiments. We then compare our results with the existing data~\cite{Conway:1989fs} and recent lattice QCD calculation for the pion and kaon PDFs~\cite{Fan:2021bcr,Salas-Chavira:2021wui}. We observed that the present numerical results for the pion and kaon PDFs are in excellent agreement with the reanalysis data~\cite{Conway:1989fs} as well as the recent lattice QCD~\cite{Fan:2021bcr,Salas-Chavira:2021wui}.

This paper is organized as follows: In Sec.~\ref{sec:nlchm}, we briefly introduce and elaborate on the formalism of the gauge-invariant nonlocal chiral-quark model (NL$\chi$QM) for the PDFs. Section~\ref{sec:result} presents our numerical results of the QDFs and GDFs for the pion and kaon with detailed discussions. Finally, the summary and conclusion are given in Sec.~\ref{sec:summary}.

\section{Nonlinear chiral quark model and parton distribution function} \label{sec:nlchm}
In this section, we briefly present a generic expression of the valence quark distribution function (QDF) for the pion and kaon. Also, we describe how to generate the gluon distribution function (GDF) for the pion and kaon via the NLO DGLAP QCD evolution. A generic expression for the twist-2 QDF for the SU(3) flavor-octet pseudoscalar (PS) meson field $\phi$ is defined by
\begin{eqnarray}
    \label{eq1}
    f_{\phi} (x) &=& \frac{i}{4\pi} \int d\eta \exp[i(xp)\cdot(\eta n)]\langle \phi (p)|\bar{q}_f (\eta n) \rlap{/}{n}q_f(0)|\phi (p)\rangle.
\end{eqnarray}
The momentum fraction of the struck quark in the PS meson is defined by $x = (k\cdot n)/(p\cdot n)$, where $n$, $k$, and $p$ are respectively the light-like vector, the parton momentum, and the PS meson momentum, respectively. Note that we have $n\cdot v = v^+$ for instance in the light-cone frame. Further details of the light-cone variables will be described in what follows.

Now we are in a position to explain the NL$\chi$QM briefly. The effective chiral action (E$\chi$A) for the NL$\chi$QM reads
\begin{eqnarray}
    \label{eq:m1}
   && \mathcal{S}^\mathrm{NL\chi QM}_{\rm{eff}} \left[\phi,m_f,\mu\right] 
=-i \mathrm{Sp}_{c,f,\gamma} \ln\left[i\rlap{/}{\partial}-\hat{m}_f  -\sqrt{M_f(i\rlap{/}{\partial})}U_5 \sqrt{M_f (i\rlap{/}{\partial})}\right],
\end{eqnarray}
where $\mathrm{Sp}_{c,f,\gamma}$ represents the functional trace over the quark color ($c$), flavor ($f$), and Lorentz indices ($\gamma$). $\hat{m}_f$ indicates the current-quark mass $\mathrm{diag}(m_u,m_d,m_s)$. In this work, we consider the isospin symmetry $m_u =m_d$ and the SU(3) flavor-symmetry breaking explicitly $m_s>m_{u,d}$. $M_f$ is the constituent quark mass for the given quark flavor and is considered a function of momentum transfer, whereas $\mu$ stands for the renormalization scale of the model. The nonlinear expression for the PS meson field $\phi$ is defined by
\begin{eqnarray}
    \label{eq:m2}
    U_5 &=& \exp \left[ \frac{i\gamma_5 \lambda \cdot \phi}{\sqrt{2}F_{\phi}}\right],\,\,\,\,
    \lambda \cdot \phi  = \left(~\begin{matrix} \frac{1}{2}\pi^0 + \frac{1}{\sqrt{6}}\eta & \pi^+ & K^+ \\ \pi^- & -\frac{1}{\sqrt{2}} \pi^0 + \frac{1}{\sqrt{6}} \eta & K^0 \\K^- & \bar{K}^0 & - \frac{2}{\sqrt{6}} \eta \end{matrix}\right),
\end{eqnarray}
where $F_\phi$ and $\lambda$ are respectively the weak-decay constants for the PS meson and the Gell-Mann matrix. 

The effective Lagrangian density for the $q$-$q$-$\phi$ interaction vertex that is obtained from the E$\chi$A is defined by
\begin{eqnarray}
    \label{eq:m3}
    \mathcal{L}^\mathrm{NL\chi QM} _{qq\phi}=\frac{i}{F_\phi} \bar{q} \Big[ \sqrt{M_f(i\rlap{/}{\partial})} 
    \gamma_5 (\lambda \cdot \phi) \sqrt{M_f (i\rlap{/}{\partial})} \Big]q.
\end{eqnarray}
As expected, by turning off the momentum dependence of the $M_f$ in Eq.~(\ref{eq:m3}), it simply turns into the well-known pseudo-scalar-type effective chiral Lagrangian density as follows:
\begin{eqnarray}
    \label{eq:m4}
    \mathcal{L}_{qq\phi}^{\mathrm{local}} = i g_{qq\phi} \bar{q} [\gamma_5 (\lambda \cdot \phi) ]q,
\end{eqnarray}
where $g_{qq\phi}$ is the $q$-$q$-$\phi$ coupling constant, which is a similar quantity obtained in the generic Nambu--Jona-Lasinio (NJL) model. For conserving the gauge invariance of the E$\chi$A in Eq.~(\ref{eq:m1}), we simply apply the minimal substitution $\partial_\mu \to D_\mu = \partial_\mu - iV_\mu$ where $V$ is the external vector field and have the following:
\begin{eqnarray}
    \label{eq:m5}
    &&\mathcal{S}^\mathrm{NL\chi QM}_{\mathrm{eff}}[\phi,m_f,V_\mu,\mu]
= -\mathrm{Sp}_{c,f,\gamma} \ln \Big[ iD\!\!\!\!/-\hat{m}_f 
    - \sqrt{M_f (iD\!\!\!\!/)}U_5 \sqrt{M_f (iD\!\!\!\!/)}\Big].
\end{eqnarray}

Using the gauge invariant E$\chi$A in Eq.~(\ref{eq:m5}), we evaluate the QDFs through a functional derivative with respect to $\phi$ and $V$, resulting in
\begin{eqnarray}
    \label{eq:m6}
\frac{\delta^3\mathcal{S}_{\mathrm{eff}}[\phi,m_f,V_\mu,\mu]}{\delta \phi^\alpha (x) \delta \phi^\beta(y)\delta V_\mu (0)}\Bigg|_{\phi^{(\alpha, \beta)},V =0},
\end{eqnarray}
where the superscripts $(\alpha, \beta)$ for the PS meson fields stand for their flavor-matrix indices. Analytically, we simply perform the expansion of the nonlinear meson field $U_5$ in the E$\chi$A up to the order of $\mathcal{O}(\phi^2)$. The expression for the QDF for the $f$-flavored quark inside the $\phi$ in the NL$\chi$QM is then obtained by
\begin{eqnarray}
    \label{eq:m7}
   f_\phi(x)&=& - \frac{iN_c}{2F_\phi^2} \int \frac{d^4k}{(2\phi)^4} \delta \left(k \cdot n - x\,p\cdot n\right) 
 \mathrm{Tr}_\gamma \Big[ \sqrt{M_b}\gamma_5 \sqrt{M_a} S_a \rlap{/}{n} S_a \sqrt{M_a}\gamma_5\sqrt{M_b} S_b \cr
    &+& \left(\sqrt{M_b}\cdot n\right) \gamma_5 \sqrt{M_a}S_a\sqrt{M_a}\gamma_5\sqrt{M_b}S_b
-\sqrt{M_b}\gamma_5 \left(\sqrt{M_a} \cdot n\right)S_a \sqrt{M_a} \gamma_5 \sqrt{M_b}S_b\Big].
\end{eqnarray}
The relevant momenta are defined by $k_a = k$ and $k_b = k-p$ in the quark propagators for the flavors $f=(a,b)$. The second and third terms of Eq.~(\ref{eq:m7}), containing ($\sqrt{M_b}\cdot n$) and ($\sqrt{M_a}\cdot n$), only appear when the momentum dependence of the effective quark mass is taken into account. These terms are so-called the \textit{nonlocal} or \textit{derivative} interaction terms that are obtained from the functional derivative of the gauge-invariant E$\chi$A with respect to the $V_\mu$. The quark propagator $S_a=S_a (k_a)$ for given flavor $a$ is expressed by
\begin{eqnarray}
    \label{eq:m8}
    S_a (k_a) \equiv \frac{\rlap{/}{k}_a+ (m_a + M_a)}{k_a^2 -(m_a+M_a)^2 + i\epsilon}
     = \frac{\rlap{/}{k}_a+ \Tilde{M}_a}{k_a^2 - M_a^2 + i\epsilon},
\end{eqnarray}
where $\Tilde{M}_a = (m_a + M_a)$ is the effective quark mass with the current quark mass $m_a$. The momentum-dependent mass functions, $M_a$ and $M_{a\mu}$, are parameterized as follows:
\begin{eqnarray}
    \label{eq:m9}
    M_a &=& M_0 \left[\frac{\mu^2}{k_a^2-\mu^2 + i\epsilon}\right]^2,\,\,\,\,
    \sqrt{M_{a\mu}} = - \frac{2\sqrt{M_a}k_{a\mu}}{(k_a^2 -\mu^2 +i\epsilon)}.
\end{eqnarray}
Note that $M_0$ is the constituent quark mass at zero momentum transfer.

Now the expression of the QDF in Eq.~(\ref{eq:m7}) can be rewritten in the light-cone coordinate using the light-cone variable which is defined by
\begin{eqnarray}
    \label{eq:m10}
    k\cdot n&=& k^+ = xP^+,\,\,\,\,k^2 = k^+ k^- - k_\perp^2,\,\,\,\,p^2 = m_\phi^2, \,\,\,\,
    k\cdot p=\frac{1}{2}(p^+ k^- + k^+ p^-).
\end{eqnarray}
Applying the above light-cone variable definitions after employing the trace in the numerator, we then arrive at the final expression for the QDF of the PS meson in the NL$\chi$QM:
\begin{eqnarray}
    \label{eq:m11}
    f_\phi(x) &=& \frac{iN_c}{4F_\phi^2} \int \frac{dk^- d^2k_\perp}{(2\pi)^3} \left[\mathcal{F}_\mathrm{L}(k^-,k_\perp^2) + \mathcal{F}_{\mathrm{NL},a} (k^-,k_\perp^2) + \mathcal{F}_{\mathrm{NL},b} (k^-,k_\perp^2)\right]+\left[x \leftrightarrow (1-x)\right].
\end{eqnarray}
where $\mathcal{F}_\mathrm{L}$, $\mathcal{F}_{\mathrm{NL},a}$, and $\mathcal{F}_{\mathrm{NL},b}$ are defined in Appexdix in detail. It is worth mentioning that QDF in Eq.~(\ref{eq:m11}) should preserve the normalization condition
\begin{eqnarray}
    \label{eq:m16}
    \int_0^1 f_\phi(x) = 1,
\end{eqnarray}
and the moments of the QDF for the PS meson can be calculated by
\begin{eqnarray}
\label{eq:m17}
    \langle x^n \rangle_{f_\phi} = \int_0^1 dx\, x^n f_\phi(x),
\end{eqnarray}
where $n=0,1,2,\cdots$ is an integer number. It is clearly seen that for $m=0$, it will reproduce the normalization condition in Eq.~(\ref{eq:m16}).

Now, we evolve the QDF to a higher factorization scale and generate the GDF using the NLO DGLAP (Dokshitzer–Gribov–Lipatov–Altarelli–Parisi) QCD evolution. First, the nonsinglet QDF distribution can be obtained by
\begin{eqnarray}
    \label{eq:m18}
    f^\mathrm{NS}_{\phi} (x) &=& f_\phi(x) - \bar{f}_\phi (x),
\end{eqnarray}
where the (anti)QDFs are respectively represented by $f_\phi(x)$ and $\bar{f}_\phi(x)$. In the DGLAP QCD evolution, the nonsinglet QDF can be generated at a higher factorization scale $Q^2$ by convoluting the splitting function as follows:
\begin{eqnarray}
\label{eq:m19}
    \frac{\partial f^\mathrm{NS}_\phi (x,Q^2)}{\partial \ln(Q^2)} &=& P_{ff}\left[x,\alpha_s(Q^2)\right] 
    \otimes f^\mathrm{NS}_\phi (x,Q^2). 
\end{eqnarray}
Here, the convolution reads
\begin{eqnarray}
    \label{eq:m20}
    P_{ff} \otimes f^{\mathrm{NS}}_\phi = \int_x^1 \frac{dz}{x} P \left[ \frac{x}{z} f^{\mathrm{NS}}_\phi (z,Q^2)\right].
\end{eqnarray}
The splitting functions can be perturbatively expanded in terms of the strong coupling $\alpha_s(Q^2)$ as $P(z, Q^2) = \sum_{n=1}\left(\frac{\alpha_s}{2\pi} \right)^nP^{(n)} (z)$. For the singlet QDF, one has
\begin{eqnarray}
    \label{eq:m21}
    f^{\mathrm{S}}_\phi (x)  = \sum_f \left[f_\phi(x) + \bar{f}_\phi(x)\right].
\end{eqnarray}
Similarly, the DGLAP evolution of the singlet QDF can be done with the GDF for the $\phi$, i.e.,  $g_\phi(x,Q^2)$ as follows:
\begin{eqnarray}
    \label{eq:m22}
    \frac{\partial}{\partial \ln(Q^2)} \Bigg[\begin{matrix}
         f^{\mathrm{S}}_\phi  (x,Q^2) \\ g_\phi(x,Q^2)
    \end{matrix} \Bigg] = \Bigg[ \begin{matrix}
        P_{ff} & P_{fg} \\
        P_{gf} & P_{gg}
    \end{matrix}\Bigg]\otimes \Bigg[ \begin{matrix}
         f^{\mathrm{S}}_\phi (x,Q^2) \\ g_\phi(x,Q^2)
    \end{matrix}\Bigg].
\end{eqnarray}
In Eq.(\ref{eq:m22}), it is clearly shown that the GDF can be obtained in the DGLAP evolution. It is worth noting that the NNLO contribution provides only negligible effects on the DGLAP evolution of the PDFs.

\section{Numerical result}\label{sec:result}
Here, we present the numerical results for the pion and kaon QDFs as well as their GDFs with detailed discussions. The constituent-quark mass is determined to satisfy the QDF normalization condition in Eq.~(\ref{eq:m16}) with the model scale $\mu =1$ GeV, resulting in $M_0=300$ MeV. Here, we use the empirical values for the PS-meson weak-decay constants as $F_\pi =$ 93.2 MeV and $F_K =$ 113.4 MeV~\cite{Nam:2012vm}. The current quark masses are chosen to be $m_u =m_d =$ 5 MeV and $m_s =$ 100 MeV. 
\begin{figure}[t]
\begin{tabular}{cc}
\includegraphics[width=8.5cm]{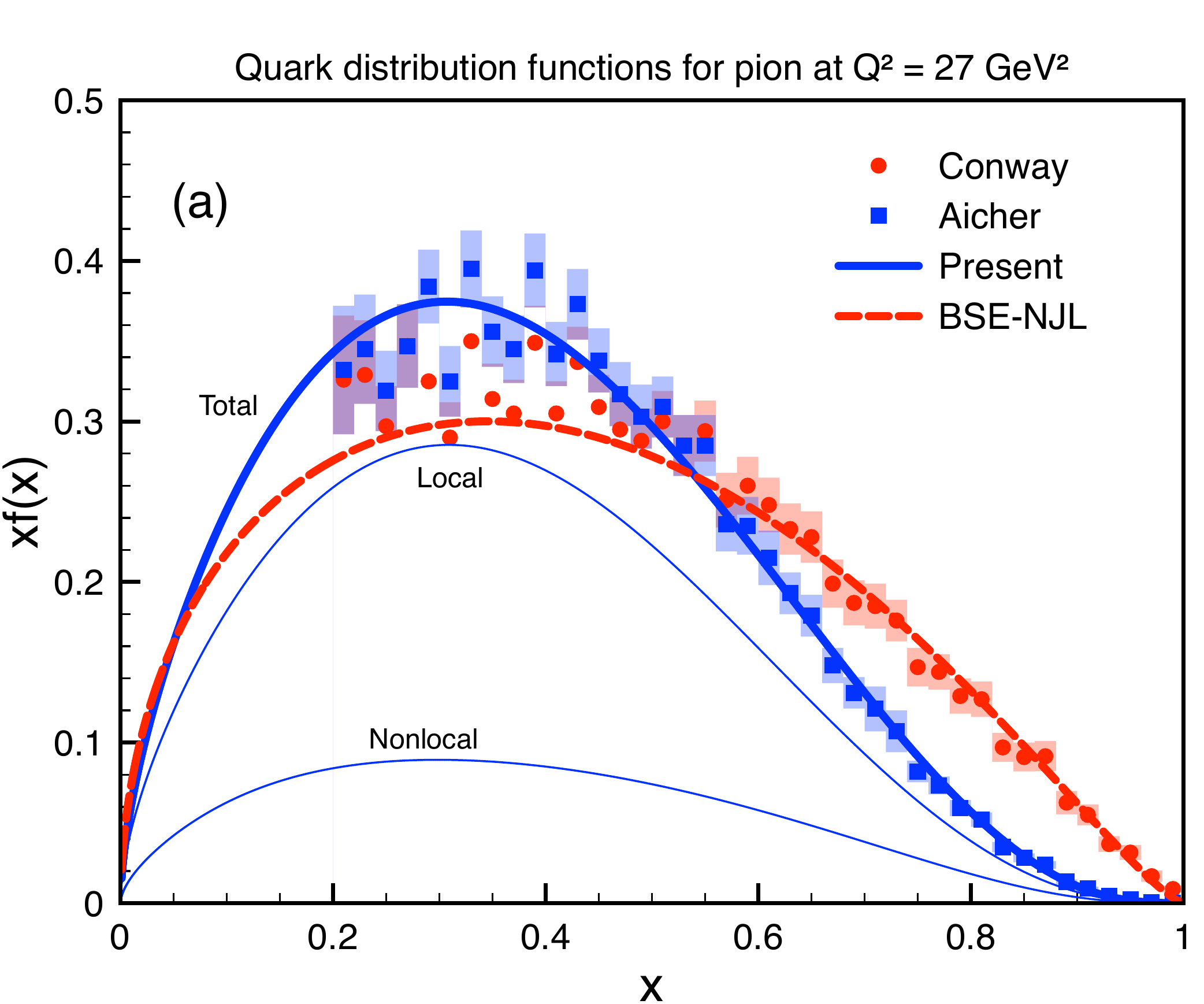}
 \includegraphics[width=8.5cm]{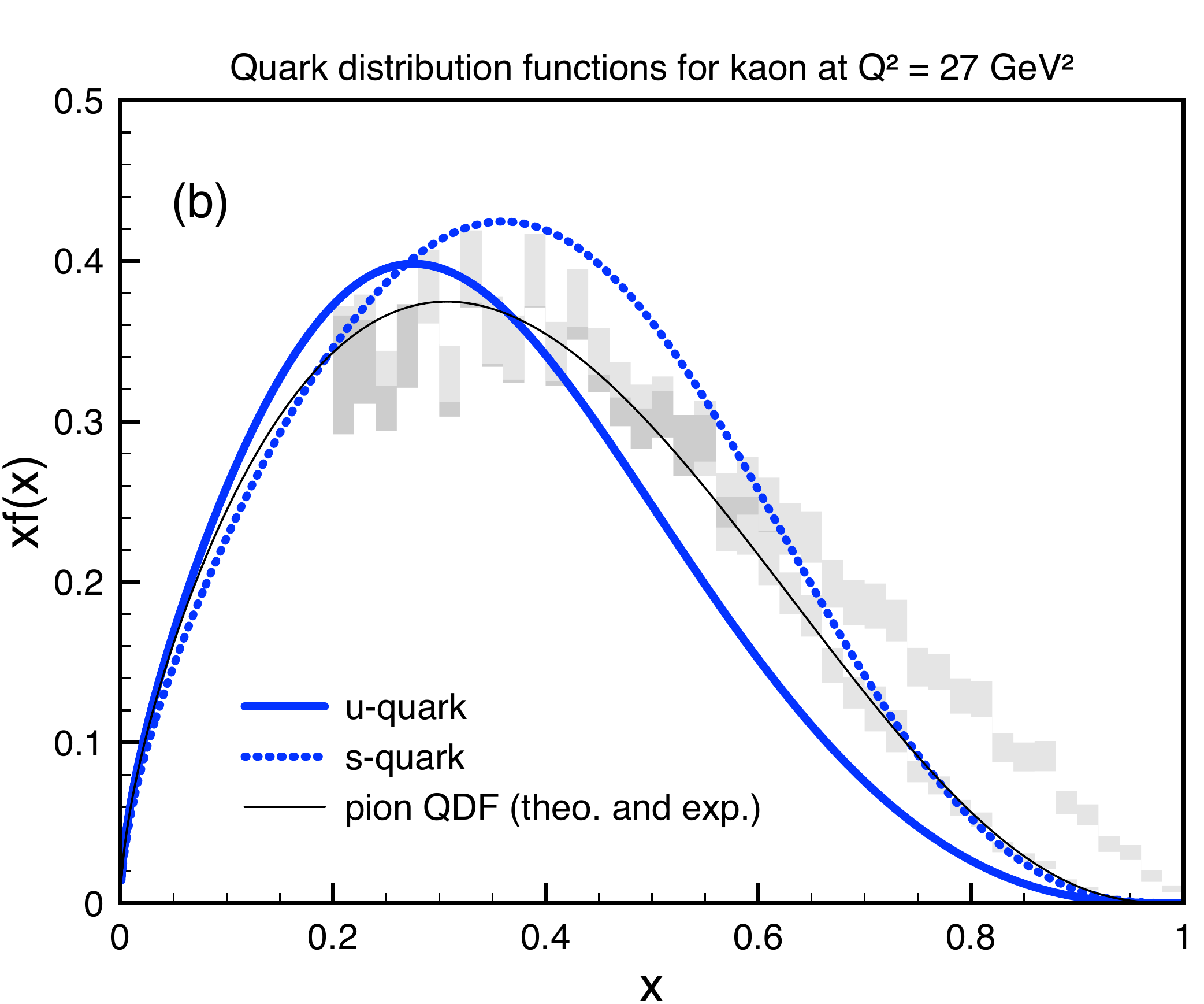}
\end{tabular}
\caption{(a) Numerical results for $xu_\pi(x)$ in NL$\chi$QM from the total (solid), local, and nonlocal (thin solid) contributions, in addition to the BSE-NJL result (dashed)~\cite{Hutauruk:2021kej}, at $Q^2=27\,\mathrm{GeV}^2$. The experimental data are taken from Refs~\cite{Conway:1989fs,Aicher:2010cb}. (b) Those for $xu_K(x)$ (solid) and $\bar{s}_K(x)$ in the same manner. For reference, $xu_\pi(x)$ in NL$\chi$QM (thin solid) and the experimental data for the pion (shades) are depicted as well.}
\label{FIG1}
\end{figure}

In panel (a) of Fig.~\ref{FIG1}, we depict the numerical results for $xu_\pi(x)$ in NL$\chi$QM from the total (solid), local, and nonlocal (thin solids) contributions, in addition to the BSE-NJL result (dashed)~\cite{Hutauruk:2021kej}, at $Q^2=27\,\mathrm{GeV}^2$ which is DGLAP evolved from the initial scale $Q_0^2 =$ 0.18 GeV$^2$. Note that the local contribution makes about $70\%$ of the total one for $xu_\pi(x)$ as usual in NL$\chi$QMl~\cite{Nam:2012vm}.  The empirical data are taken from Refs~\cite{Conway:1989fs,Aicher:2010cb}. It is worth mentioning that the curve of Ref.~\cite{Aicher:2010cb} (square) is the new reanalysis of the data of Ref.~\cite{Conway:1989fs} (circle). We note that the reanalyzed empirical data are reproduced excellently via NL$\chi$QM, whereas the BSE-NJL model fits well with the old curve of Ref.~\cite{Conway:1989fs}. Considering that the momentum dependencies of the quark interactions are properly taken into account in NL$\chi$QM, we can conclude that the reanalyzed data is more reliable indicating the relevant physics corresponding to the nontrivial quark interactions in the instanton vacuum for instance~\cite{Nam:2012vm}. The end-point behaviors of the two theoretical models are quite distinguishable in the vicinity of $x=1$. This observation may provide an explanation for the long-standing puzzle of the power-law behavior of QDF at $x\to1$.

 In order to see the large-$x$ behavior of QDFs clearly, we fit the numerical result of NL$\chi$QM for $x=[0.8,1.0]$ using the power-law form $(1-x)^{r}$, where $r$ denotes a real number. By doing this, we find that $xu_\pi(x)$ has the power-law behavior of $(1-x)^{1.85}$ in the vicinity of $x=1$. This observation is consistent with the theoretical result of Ref.~\cite{Cui:2021mom}, which also considers the momentum-dependent interactions. Similarly, the power-law behavior for the BSE-NJL model is given by $(1-x)^{1.23}$. For practical purpose, we present the parameterizations of $xu_\pi(x)$ at a factorization scale as follows:
\begin{eqnarray}
xu_\pi (x) &=&  x^{0.60} (1-x^{2.00})^{2.74}\,\,\mathrm{at}\,\,Q^2 = 27\,\mathrm{GeV}^2.
\end{eqnarray}

In panel (b) of Fig.~\ref{FIG1}, we show the numerical results for the kaon, i.e.,  $xu_K(x)$ (solid) and $\bar{s}_K(x)$ (dotted) in the same manner with that for the pion case. For reference and comparison, $xu_\pi(x)$ in NL$\chi$QM (thin solid) and the experimental data for the pion (shades) are given as well. Because of the considerable mass difference between the light and strange quarks inside the kaon, it is obvious that the peak positions of the kaon QDFs deviate from that for the pion. The peak position for the strange quark is shifted to the larger momentum fraction and this behavior can be well understood because the heavier one carries more momentum than the light one. The power-law behavior for the $xu_K(x)$ is given by $(1-x)^{2.30}$, being different from that of $xu_\pi(x)$. Similarly to the pion case, we also provide the parameterizations for the kaon QDFs for practical purposes as follows:
\begin{eqnarray}
xu_K (x) = x^{0.59} (1-x^{2.58})^{5.18},\,\,\,\,x\bar{s}_K (x) = x^{0.64} (1-x^{2.86})^{3.87}\,\,\mathrm{at}\,\,Q^2 = 27\,\mathrm{GeV}^2.
\end{eqnarray}

We also compute the various moments of QDFs at $Q^2 =$ 27 GeV$^2$ for the pion and kaon are summarized in Table~\ref{tab1}. As shown in the table, the quark inside the PS meson carries about $20\%$ of the longitudinal momentum of the meson by seeing the first moments $(n=1)$.
\begin{table}[h]
\begin{ruledtabular}
\renewcommand{\arraystretch}{1.3}
\caption{Various moments for the pion and kaon QDFs at $Q^2 =$ 27 GeV$^2$.}
\label{tab1}
\begin{tabular}{ccccccc}
&  $n=1$ & $n=2$  & $n=3$ & $n=4$ & $n=5$ & $n=6$  \\ \hline
$\langle x^n \rangle_{u_\pi}$ &  0.207 & 0.077 & 0.037 & 0.020 & 0.012 & 0.008 \\
$\langle x^n \rangle_{u_K}$  & 0.189 & 0.064 & 0.027 & 0.014 & 0.007 & 0.004 \\
$\langle x^n \rangle_{\bar{s}_K}$  & 0.229 & 0.088 & 0.042 & 0.023 & 0.013 & 0.008 \\
\end{tabular}
\end{ruledtabular}
\end{table}

\begin{figure}[h]
\begin{tabular}{cc}
\includegraphics[width=8.5cm]{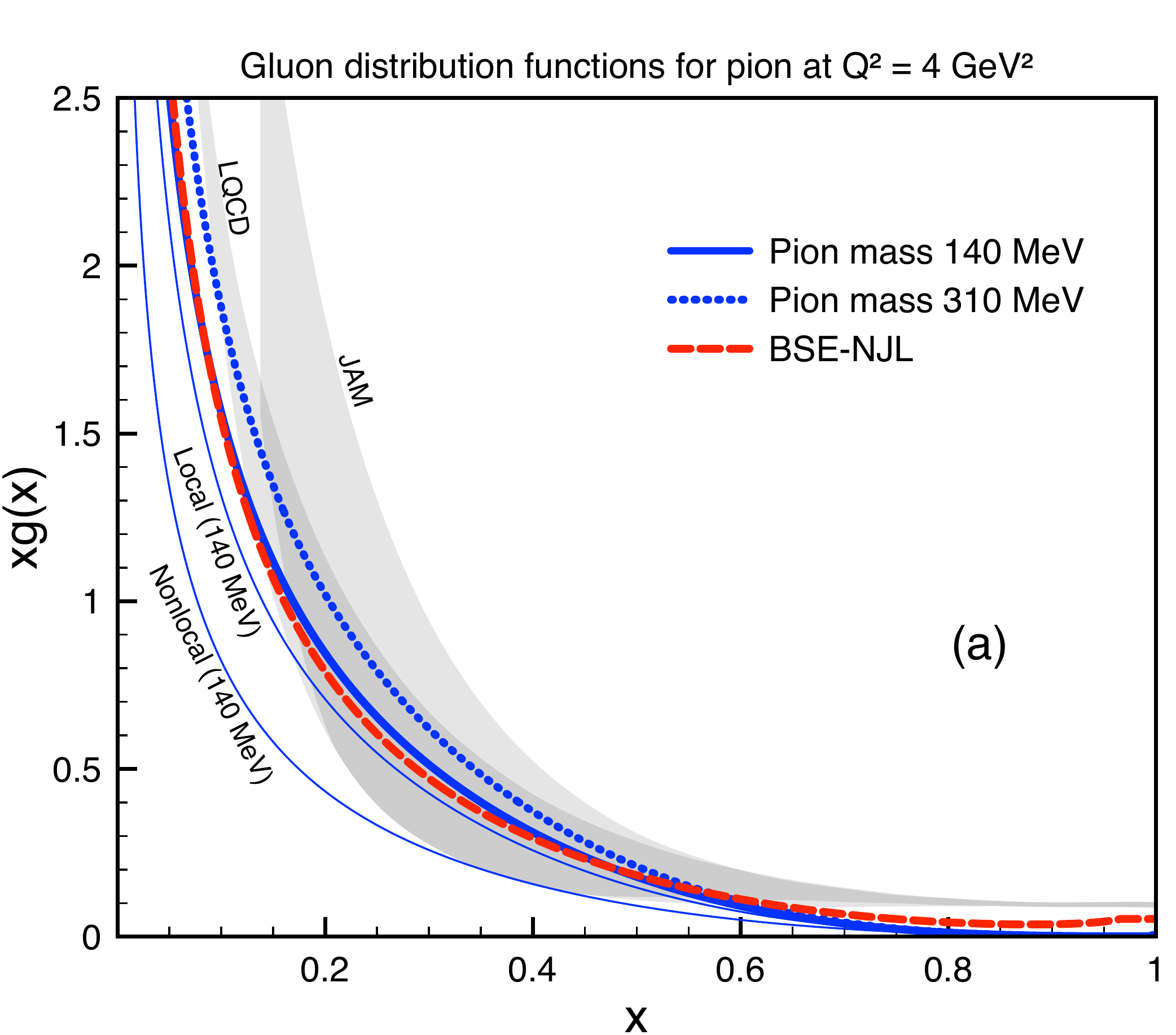}
\includegraphics[width=8.5cm]{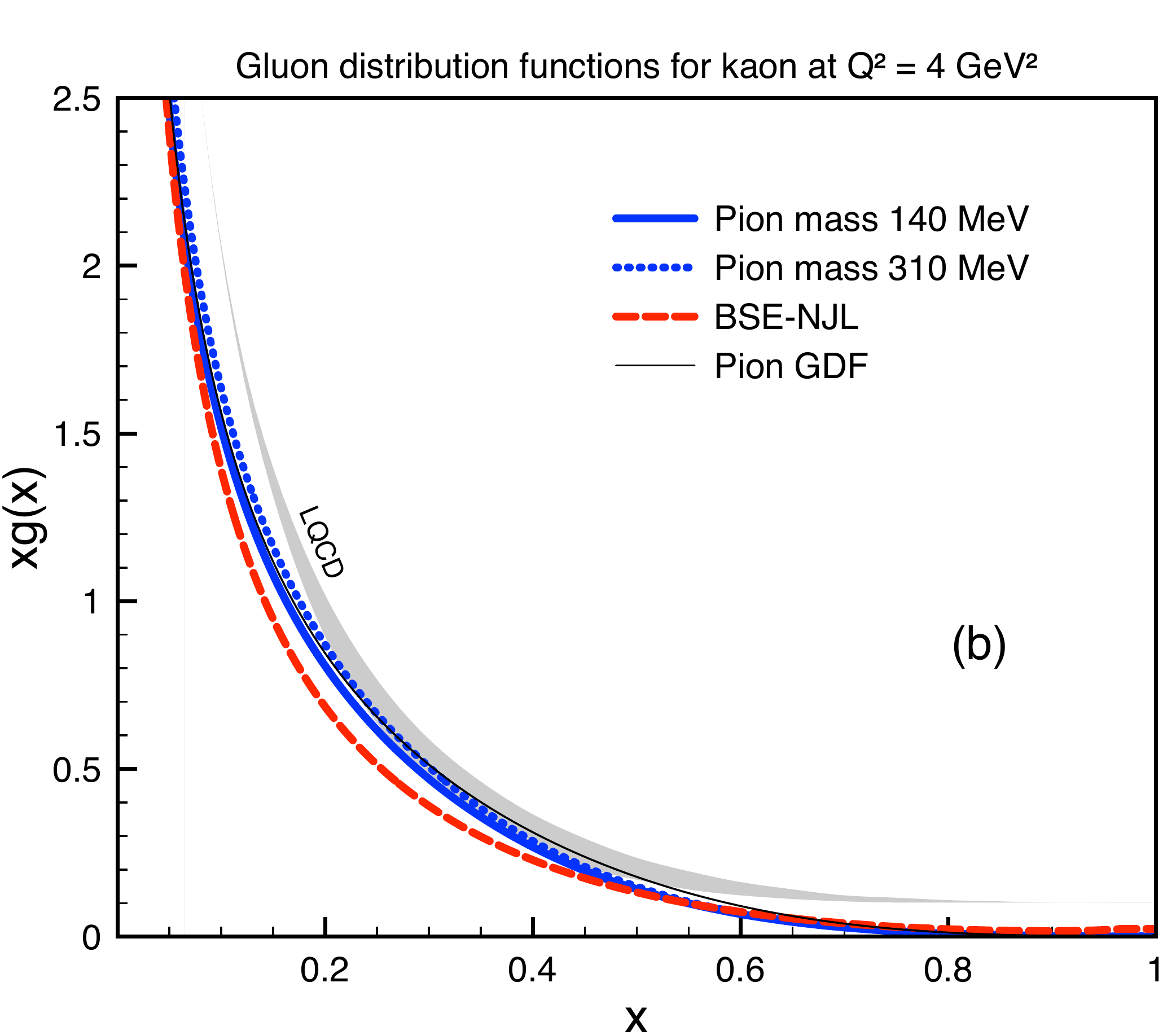}
\end{tabular}
\caption{(a) Numerical results for $xg_\pi(x)$ in NL$\chi$QM from the total (solid), local, and nonlocal (thin solids) contributions with $m_\pi=140$ MeV at $Q^2=4\,\mathrm{GeV}^2$. We also show the results of the present model with $m_\pi=310$ MeV (dotted) and BSE-NJL one (dashed). The shaded areas represent the JAM global analysis~\cite{Barry:2021osv} and lattice QCD data~\cite{Salas-Chavira:2021wui}. (b) Those for $xg_K(x)$ in the same manner. The thin solid line denotes the pion GDF for reference.}
\label{FIG2}
\end{figure}
Now, we are in a position to provide the numerical results for the pion and kaon GDFs in NL$\chi$QM at $Q^2=4\,\mathrm{GeV}^2$. In the panel of Fig.~\ref{FIG2}, we present those for $xg_\pi(x)$ with $m_\pi=140$ MeV (solid) and $310$ MeV (dotted). The later value of $m_\pi$ corresponds to that considered in the lattice-QCD (LQCD) simulation data~\cite{Salas-Chavira:2021wui} (shade). We also show the results from the BSE-NJL model for comparison (dashed). The fitting curve from the Jefferson-Lab Angular Momentum (JAM) global analysis is depicted in shade as well~\cite{Barry:2021osv}. We observe a similar tendency for the local and non-local contributions in NL$\chi$QM to that for QDF. Note that the two theory curves are almost consistent showing small deviations in the region of $x=0.2$ and $x=1$. As the pion mass increases, the NL$\chi$QM curve approaches the LQCD data as expected. Qualitatively, the present result from NL$\chi$QM reproduces the LQCD and JAM analysis for $x\gtrsim0.2$ as shown in the figure. As for $x\lesssim0.2$, one needs more statistics from experiments to pin down the correct behavior of GDF. 

In panel (b) of Fig.~\ref{FIG2}, the numerical results for $xg_K(x)$ are given in the same manner as panel (a). For reference and comparison, we also draw the curve for $xg_\pi(x)$ (thin-solid). Being different from QDF, GDF does not show significant differences between the pion and kaon cases, since the gluons are blind to the quark flavors. Moreover, the pion-mass dependence is considerably weak, since there is a heavier mass scale $m_s=100$ MeV for the kaon case. Interestingly, the difference between the BSE-NJL model and NL$\chi$QM results becomes more obvious than the pion case, and the latter one qualitatively reproduces the LQCD data~\cite{Salas-Chavira:2021wui} better for $x\gtrsim0.2$. For a better understanding of the kaon GDF, future experiments such as the EIC, EicC, and COMPASS++/AMBER are really required to confront these theoretical results. The power law of $xg_\pi(x)$ behaves as $(1-x)^{3.06}$ at the large-$x$ region, whereas $(1-x)^{3.17}$ for $xg_K(x)$ at $Q^2=4\,\mathrm{GeV}^2$. Similarly to the QDFs, we parametrize the GDFs for the pion and kaon as follows: 
\begin{eqnarray}
    \label{eq:nr1}
 xg_\pi (x) = 1.49x^{-0.35}(1-2.06x^{0.5}+3.46x)(1-x)^{3.88} \,\,\,\,\mathrm{at}\,\,Q^2 = 4\,\mathrm{GeV}^2, \\ 
 xg_K (x) = 1.48x^{-0.35}(1-2.13x^{0.5}+3.97x)(1-x)^{4.45} \,\,\,\mathrm{at}\,\,Q^2 = 4\,\mathrm{GeV}^2.
\end{eqnarray}

The moments of the pion and kaon GDFs are listed in Table~\ref{tab2}. From the table, it is found that the gluon carries about $60\%$ of the longitudinal momentum of the PS meson. 
\begin{table}[h]
	\begin{ruledtabular}
		\renewcommand{\arraystretch}{1.3}
		\caption{Various moments for GDFs for the pion and kaon at $Q^2 = 4\,\mathrm{GeV}^2$.}
		\label{tab2}
		\begin{tabular}{ccccccc}
		&  $n=1$ & $n=2$  & $n=3$ & $n=4$ & $n=5$ & $n=6$  \\ \hline
        $\langle x^n \rangle_{g_\pi}$ &  0.605 & 0.079 & 0.023 & 0.009 & 0.004 & 0.0022 \\
		         $\langle x^n \rangle_{g_K}$ & 0.577 & 0.071 & 0.019 & 0.007 & 0.003 & 0.002 \\ 
		\end{tabular}
	\end{ruledtabular}
\end{table}

\begin{figure}
\includegraphics[width=8.5cm]{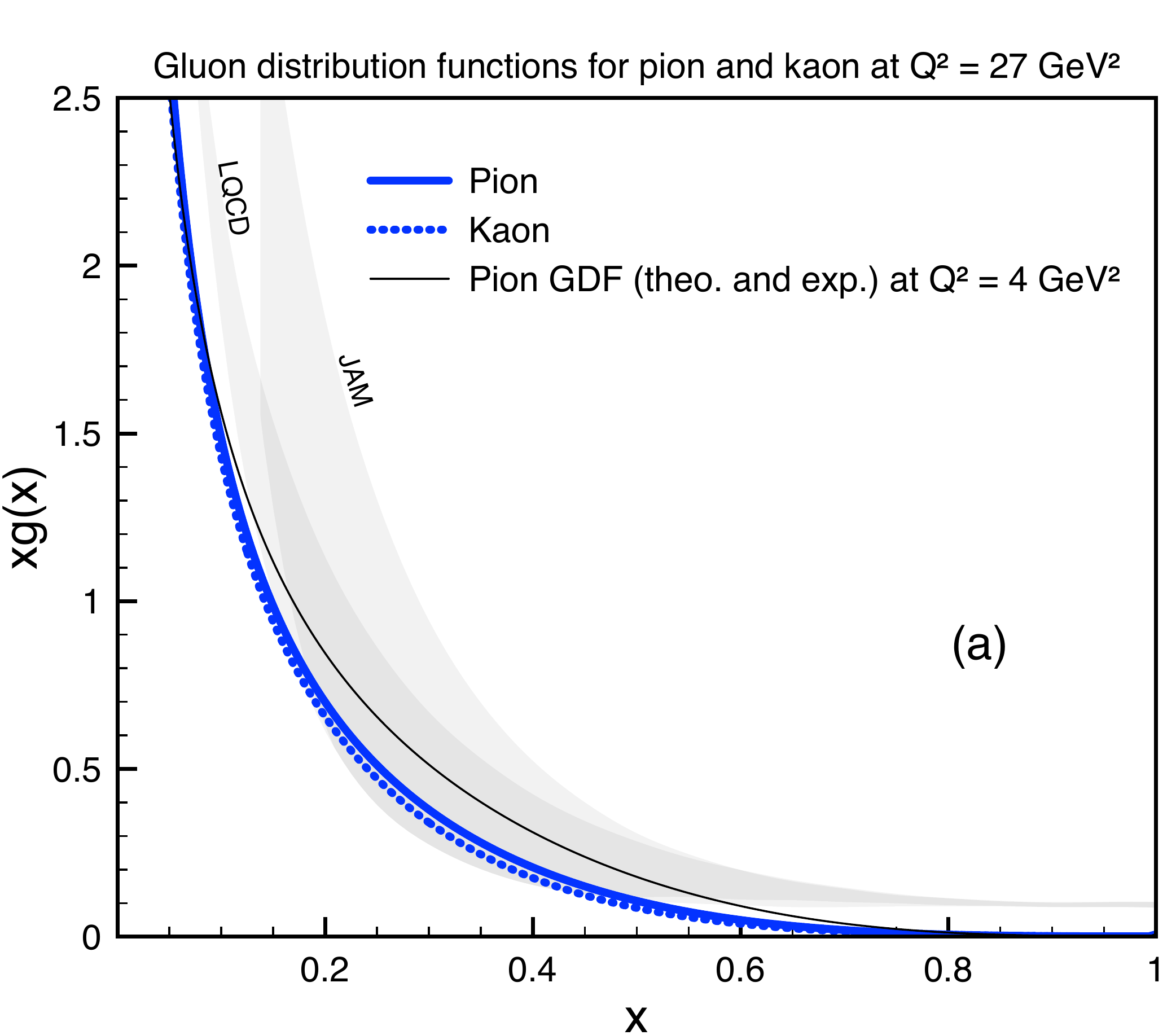}
\includegraphics[width=8.5cm]{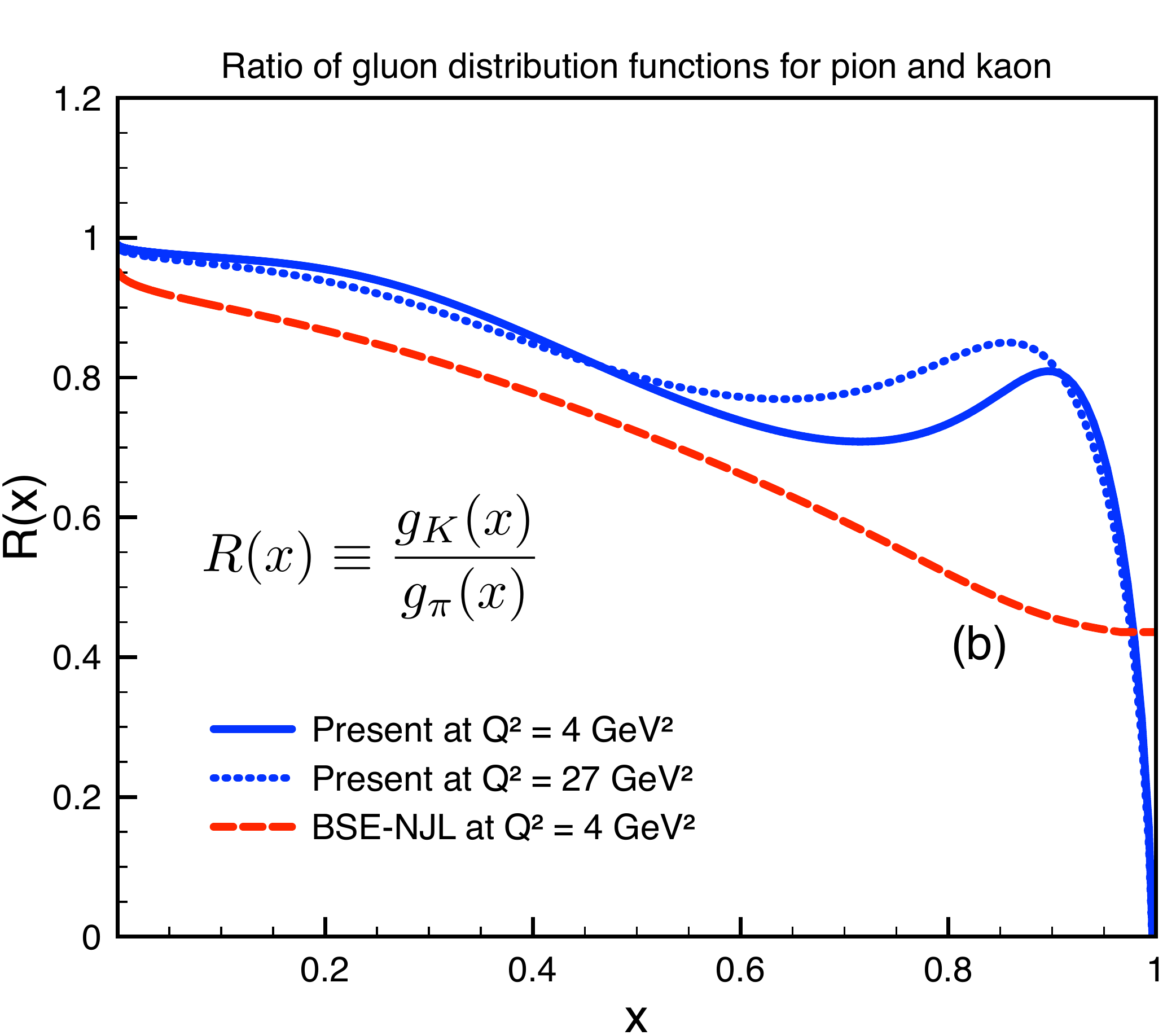}
\caption{(a) Numerical results for $xg_\pi(x)$ (solid) and $xg_K(x)$ (dotted), evolved to $Q^2=27\,\mathrm{GeV}^2$. The thin-solid line and shaded areas stand for the pion GDF including the lattice QCD and JAM data at $Q^2=4\,\mathrm{GeV}^2$ for reference. (b) The ratio $R(x)\equiv{g_K(x)}/{g_\pi(x)}$ at $Q^2=4\,\mathrm{GeV}^2$ (solid) and $27\,\mathrm{GeV}^2$ (dotted). The result of the BSE-NJL model at $Q^2=4\,\mathrm{GeV}^2$ is shown as well (dashed).}
\label{FIG3}
\end{figure}

Next, we investigate GDFs for the pion (solid) and kaon (dotted) evolved to the factorization scale at $Q^2=27\,\mathrm{GeV}^2$, shown in panel (a) of Fig.~\ref{FIG3}. For comparison, we also show the LQCD data and JAM global analysis for the pion case in shades, in addition to the pion at GDF $Q^2=4\,\mathrm{GeV}^2$ (thin solid). It turns out that, as the factorization scale gets larger, the GDFs decrease more stiffly as functions of $x$, i.e., manifesting the weaker nonperturbative gluon contributions. To verify the $Q^2$-dependent behavior of GDF, one needs more experimental data.  

Finally, we compute the ratio of the pion and kaon GDFs which is defined by
\begin{equation}
\label{eq:RR}
R(x)=\frac{g_K(x)}{g_\pi(x)},
\end{equation}
in order to see the difference between them quantitatively. The numerical results are depicted in panel (b) of Fig.~\ref{FIG3}. The NL$\chi$QM curves are given for $Q^2=4\,\mathrm{GeV}^2$ (solid) and $Q^2=27\,\mathrm{GeV}^2$ (dotted). The difference between the PS mesons becomes obvious as $x$ increases over $x\approx2$, whereas the difference depending on the $Q^2$ values is marginal. Since the gluon dynamics dominate at the small $x$ region, the flavor dependence of the PS mesons does not make a significant difference there as shown in the figure. As the $x$ increases, the quark dynamics in addition to the gluon contribution inside the PS mesons becomes significant, resulting in the smaller $R(x)$ than unity. Interestingly, there appear bump structures at $x\approx0.8$ and it can be understood by the delicate interplay between the gluon and quark contributions in the DGLAP evolution of the singlet PDFs. On the contrary, the BSE-NJL model curve does not show the bump and decreases monotonically as a function of $x$. In other words, the momentum-dependent quark interaction, which is manifested in NL$\chi$QM, is crucial to produce the bump structure in $R(x)$.

\section{Summary} \label{sec:summary}

In the present work, we have investigated the gluon distribution functions (GDF; $g_\phi$) as well as the quark ones (QDF; $f_\phi$) for the PS meson ($\phi=\pi,K$) in the framework of the gauge-invariant nonlocal chiral-quark model (NL$\chi$QM), which properly considers the momentum dependence of the quark interactions. We then dynamically generate the GDF  \textit{via} the splitting functions in the NLO DGLAP QCD evolution using the singlet QDF. 

We find that the numerical result for $xg_\pi(x)$ at $Q^2 = $ 4 GeV$^2$ shows an excellent agreement with the recent lattice-QCD data~\cite{Fan:2021bcr} as well as JAM global analyses~\cite{Barry:2021osv}. This satisfactory description of the data is followed by the fact that the numerical result for $xu_\pi(x)$ at $Q^2 =$ 27 GeV$^2$ which is used to generate $xg_\pi(x)$ reproduces the reanalyzed experimental data~\cite{Aicher:2010cb} qualitatively very well for the wide range of $x$. Note that $xu_\pi(x)$ from the BSE-NJL model without momentum-dependent interactions fails to describe the reanalyzed data, whereas it is consistent with the old experimental data~\cite{Conway:1989fs}. These results may provide good reasoning for the long-standing puzzle of the power-law behavior at $x \approx 1$.

We also provide the numerical results for the kaon, i.e., $xu_K(x)$ and $x\bar{s}_K(x)$ at $Q^2=4~\mathrm{GeV}^2$ and $27~\mathrm{GeV}^2$, and they are consistent with the recent lattice data within the errors~\cite{Salas-Chavira:2021wui}. Although no data are available for the kaon at the moment, the present results for the kaon will be useful for possible future experiments. Similarly to the QDFs from the NL$\chi$QM, the local and nonlocal contributions,  and kaon, where the nonlocal effect is small but contributes significantly to reproducing the data.

Results for ratios of the gluon PDFs in the kaon and in the pion at $Q^2 =$ 4 GeV$^2$ show that the gluon PDFs in the pion are larger than that in the kaon, which is consistent with the Dyson-Schwinger equation (DSE) result~\cite{Cui:2021mom}, which also considers momentum dependent in the model. However, the ratio a bit increase around $x \simeq$ 0.9 and it then decreases again. It is expected due to the transition region from the soft to hard scales, as also found in the DSE model, where such behavior could not be found in the BSE-NJL model. Overall, one can conclude that the gluon PDFs for the pion are larger than that in the kaon, which is consistent with other theoretical findings. Furthermore, for practical purposes, we also do a parameterization for pion and kaon's gluon and quark PDFs, which are useful for other calculations.

Our findings, in the present work, on the gluon distribution functions for the pion and kaon are really needed to be confronted by future modern facilities of the electron-ion colliders (EIC)~\cite{Arrington:2021biu}, electron-ion colliders in China (EicC)~\cite{Anderle:2021wcy}, and AMBER-SPS COMPASS~\cite{Adams:2018pwt} experiments. Also, our results for the quark and gluon PDFs with local and nonlocal contributions would be interesting guidance and information for the lattice QCD.

\section*{Acknowledgements}
P.T.P.H. thanks Huey-Wen Lin (Michigan State University) for providing us with their recent lattice QCD calculation results for the gluon distribution functions for the pion and kaon. This work was supported by the National Research Foundation of Korea (NRF) grant funded by the Korean government (MSIT) (No.~2022R1A2C1003964). The work of S.i.N. is also partially supported by the NRF grants funded by the MSIT (No.~2018R1A5A1025563 and No.~2022K2A9A1A0609176).

\section*{Appendix}
The relevant functions in Eq.~(\ref{eq:m11}), i.e., $\mathcal{F}_\mathrm{L} (k^-,k_\perp^2)$, $\mathcal{F}_{\mathrm{NL},a} (k^-,k_\perp^2)$, and $\mathcal{F}_{\mathrm{NL},b} (k^-,k_\perp^2)$ are defined as follows:
\begin{eqnarray}
    \label{eq:m12}
    \mathcal{F}_{L} (k^-,k_{\perp}^2) &=& \frac{4p^+ \eta^2 D_b^4 \Big[\mathcal{N}_1 + \mathcal{N}_2 + \mathcal{N}_3 \Big]}{\mathcal{D}_1 \mathcal{D}_2},\,\,\,\,
    \mathcal{F}_{\mathrm{NL},a} (k^-,k_{\perp}^2) = \frac{4p^+ \eta^2 \Big[\mathcal{N}_4 + \mathcal{N}_5 + \mathcal{N}_6 \Big]}{\mathcal{D}_1 \mathcal{D}_2}\Big[ \frac{x}{D_a^2}\Big],\cr
    \mathcal{F}_{\mathrm{NL},b} (k^-,k_{\perp}^2) &=& \frac{4p^+ \eta^2 \Big[\mathcal{N}_4 + \mathcal{N}_5 + \mathcal{N}_6 \Big]}{\mathcal{D}_1 \mathcal{D}_2} \Big[ \frac{x}{D_b^2}\Big],
\end{eqnarray}
where the $\mathcal{N}_{1\sim6}$ are given by 
\begin{eqnarray}
\label{eq:m13}
    \mathcal{N}_1 &=& \left[ D_a^4 D_b^8((2-x)k_{\perp}^2 + (1-x)^2k^-p^+)\right], \,\,\,\,
    \mathcal{N}_2 = \left[ 2(1-x)D_b^4(D_b^4m_b+\eta)(D_a^4m_a+ \eta)\right], \cr
    \mathcal{N}_3 &=& \left[xD_a^4 (D_b^4m_b + \eta)^2 \right],\,\,\,\,
    \mathcal{N}_4 = \left[ \eta (D_b^4m_b +\eta) \right],\,\,\,\,\mathcal{N}_5 = \left[ D_a^4(m_a \eta)\right],
    \cr
    \mathcal{N}_6 &=& \Big[ D_a^4D_b^4(2k_\perp^2 + (1-2x)k^-p^+ + m_a m_b+ x m_\phi^2)\Big],
\end{eqnarray}
and $\mathcal{D}_{1,2}$ read 
\begin{eqnarray}
\label{eq:m14}
    \mathcal{D}_1 &=& \left[ D_a^8 (\zeta_a -\alpha k^-) + 2m_a\eta D_a^4+ \eta^2 \right]_a,\,\,\,\,
    \mathcal{D}_2= \left[ D_b^8(\zeta_b-\beta k^- + \delta)+2m_b \eta D_b^4 + \eta^2 \right]_b^2.
\end{eqnarray}
We also introduce the following notations and expressions for simplicity:
\begin{eqnarray}
    \label{eq:m15}
    \alpha &=& xp^+,\,\,\,\,\beta = -(1-x)p^+,\,\,\,\,\gamma = k_\perp^2 + \mu^2, \,\,\,\,
    \delta = -(1-x)m_\phi^2,\,\,\,\,\eta = M_0\mu^4, \cr
    \zeta_a &=& k_\perp^2 + m_a^2,\,\,\,\,\zeta_b= k_\perp^2 + m_b^2, \,\,\,\,
    D_a^2 = \gamma -\alpha k^- +\delta,\,\,\,\,,D_b^2 =\gamma-\beta k^-,\,\,\,\,
    p^2 = m_\phi^2 = p^+ p^- - p_\perp^2.
\end{eqnarray}

\end{document}